# MULTI-LEVEL SELECTION IN BIOLOGY

Author: George F R Ellis

Address: Mathematics Department and ACGC, University of Cape Town;
Trinity College and DAMTP, Cambridge University.

**Abstract:** Okasha (2006) proposed distinguishing aspects of selection: those based in particle level traits (MSL1), and those based in group level traits (MSL2). It is proposed here that MSL1 can usefully be further split into two aspects, one ($MLS1_E$) representing selection of particles based in their individual interaction with environmental properties, and one ($MLS1_G$) representing Multi Level Selection of particles based in their relation to group properties. Similarly MSL2 can be split into two parts based in this distinction. This splitting enables a characterisation of how emergent group properties can affect particle selection, and thus affect group traits that are important for survival. This proposal is illustrated by considering a key aspect of animal and human life, namely the formation of social groups, which greatly enhances survival prospects. The biological mechanism that underlies such group formation is the existence of innate primary emotional systems studied by Panksepp (1998), effective through the ascending systems in the brain that diffuse neurotransmitters such as dopamine and norepinephrine through the cortex. Evolutionary emergence of such brain mechanisms, and hence the emergence of social groups, can only result from multi-level selection characterised by the combination of MSL2 and $MLS1_G$. The distinctions proposed here should be useful in other contexts.

**Keywords**: Selection, Multi-level, Emergence

## 1: INTRODUCTION

Samir Okasha's book *Evolution and the Levels of Selection* (Okasha 2006) gives a comprehensive discussion of the ongoing debate about levels of selection in evolutionary biology. Massimo Pigliucci's review (Pigliucci 2010) emphasizes that an important feature of the book is the distinction made therein between Multi Level Selection-1 (MSL1), concerned with the evolution of particle-level traits, and Multi Level Selection-2 (MLS2), concerned with the collectives themselves and their evolution. This distinction helps considerably in clarifying some of the disputes that have arisen as regards multilevel selection processes.

The new suggestion made in this note is that it may be useful to refine Okasha's proposal by defining $MLS1_E$, selection of individuals due to the environmental context independent of the existence of the group, leading to group **fitness-1**, and $MLS1_G$, selection of individuals that is essentially due to the existence of the group as an emergent entity, leading to group **fitness-2**. These definitions at the individual level will be reflected by corresponding definitions at the group level: $MLS2_E$ is the group level selection effect due to aggregation of individual advantage characterised by $MLS1_E$. In contrast to this, $MLS2_G$ is group level selection that occurs specifically because of the existence of the group as an emergent entity, resulting in properties at the individual level being selected for by $MLS1_G$. The combination of $MLS2_G$ and $MLS1_G$ is a form of selection of individuals that is essentially multilevel in character.

In order to make the case that this proposal of distinguishing $MLS1_G$ and $MLS1_E$ can work and is useful, it is necessary to give an example where this distinction can be convincingly made. That is the topic of Section 2, discussing group formation in all higher animals (including humans). It is then necessary to show there is some payoff from this proposal. Okasha discusses how a linear regression model is often better than Price's Equation when investigating multilevel selection. Section 3 considers how the proposal made here will be reflected in a modified version of that linear regression model, making the multi-level relations clearer. Section 4 presents an underlying biological mechanism (innate emotional systems) which plausibly explains animal group formation as discussed in Section 2. This mechanism is realised through specific aspects of the brain, namely the `ascending systems' linking nuclei in the limbic system to the cortex. It is suggested that existence of this biological mechanism and the associated brain structures can only be explained by multilevel selection, as considered in the preceding sections. Finally Section 5 briefly considers how this proposal fits into the bigger picture of evolutionary theory.

**2: EVOLUTION AND THE ISSUE OF GROUP FORMATION**

This section considers the core issue of group formation in all higher animals. To make the situation specific, consider the case of why animals, say buffalo in the Kruger National Park, or people, say Bushmen in the Kalahari, find it important to group together in herds or tribes. The key point is that a collection of buffalo wandering around on their own is not a herd; it is an aggregation (the whole is just the sum of its parts). If the same buffalo band together as a herd interacting in a social way, they can collectively protect each other, sensing danger earlier than when alone and acting together to protect each other, and so are far more likely to survive. A specific example: a herd can act together as a unit with the common purpose of warding off danger of attack by lions.[1] The same is true for a group of Bushmen: living isolated lives on their own, they are vulnerable to many dangers and will find it difficult to get food. If they band together, on the other hand, they can act collectively to protect young, detect dangers, ward off predators, and hunt together, and can share food, skills, resources, and information. The whole is much more than the sum of its parts; the young are much more likely to survive. A similar example is social learning in primate foraging decisions (de Waal,

---

[1] For a graphic demonstration of the protection provided in this context by the common purpose of a collective of buffalo, see http://www.youtube.com/watch?v=xHIkUzRw2jw.

Borgeaud, and White 2013). It is crucial that culture and technology can evolve in this context. A specific example: the skills of animal tracking (Liebenberg 2001)[2] are passed down from generation to generation through a group educational process. This collective process, impossible unless the group acts as a collective, greatly enhances the survival prospects of the group. Banding together into social groups probably played a key role in the evolution of language and intelligence (Donald 1991, Pringle 2013), which enabled human domination over all other species.

The selection dynamic in action is shown in diagram 1.

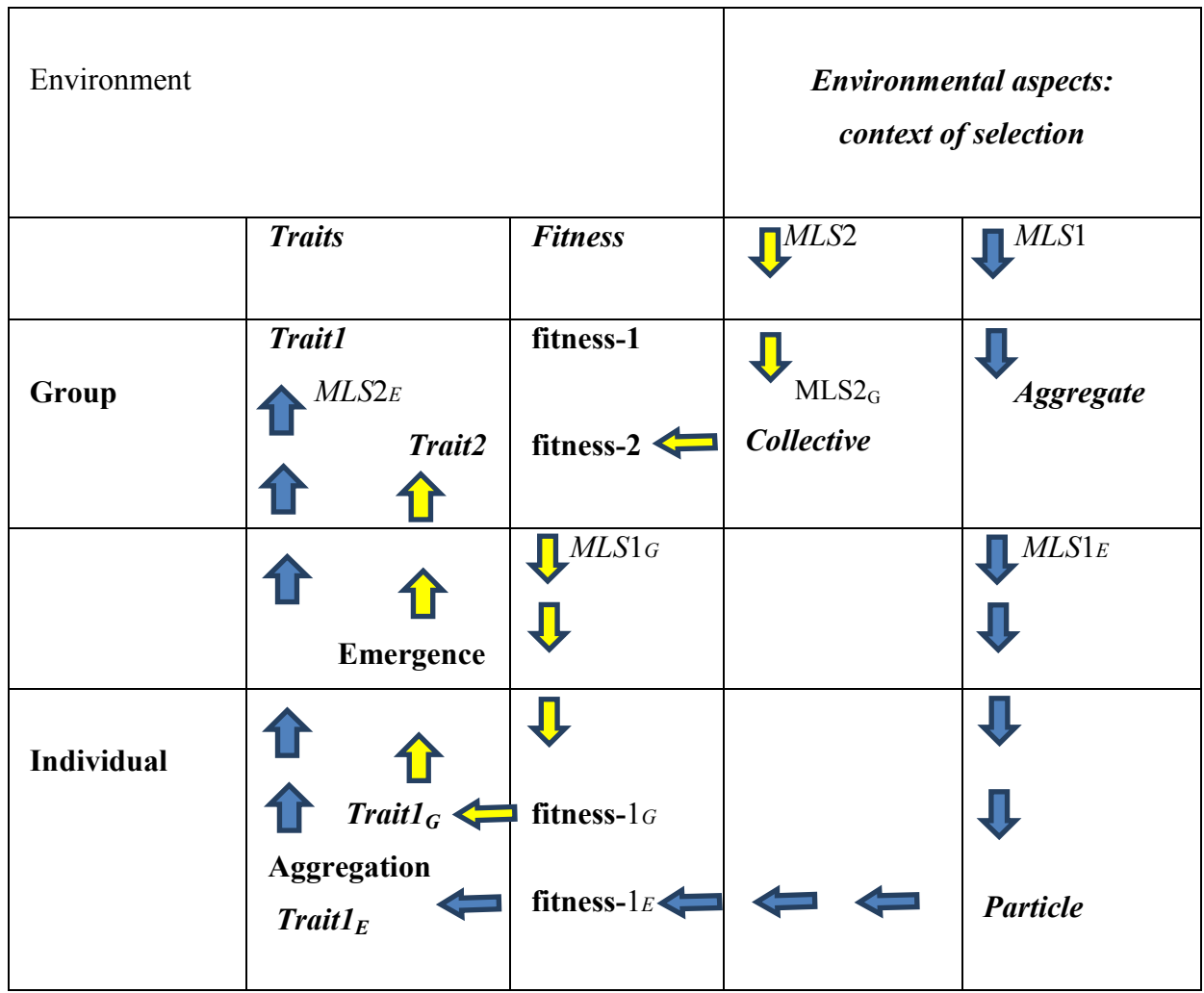

**Diagram 1**: *Multilevel selection: selection effects between the environment, the collective, and the individual. MLS1$_E$ acts on individuals irrespective of the group but MLS1$_G$ acts on individuals because of the existence of the group. Consequently MLS2$_E$ emerges as a group level selection effect due to aggregation. By contrast MLS2$_G$ is group level selection essentially due to group existence.*

---

[2] See http://www.cybertracker.org/tracking/basics-of-tracking/spoor-identification for some of its aspects.

The individual level Trait1$_E$ is an individual trait that gives an advantage to the individual in the context of the environment, and this advantage has nothing to do with the existence of the group. Examples are individuals being able to run very fast, being very strong, and so on. Selection for such traits is thus the case of selection based on individual traits alone, and can operate in an unstructured population in which there are no groups at all, but it will also operate when groups exist. This trait confers individual fitness-1$_E$ in regard to the overall environment, and is selected for by MLS1$_E$.

It is not multi-level selection as regards the individuals, in that it is just selection based on properties of the individual. However aggregation of the individual Trait1*E* over all the members of the group improves group survival capacity, and so underlies the group level Trait1 (the group is more likely to survive if it is made of stronger and faster members). Even though this is "nothing but" the sum of the parts, it enhances group survival. This leads to multilevel selection MLS2$_E$ at the group level, in that this aspect of the group's fitness is due to aggregation of the enhanced fitness of its members.

Trait2 is a group trait that gives a selective advantage to the group as a whole *because the group is an emergent entity*, acting as a collective. In the case of the buffalo, examples are the herd moving together to seek water in places remembered by older animals, forming a defensive ring against predators, and so on. These obviously can't be a trait of the individuals (one buffalo is not a herd, which by will definition require at least 5 members; less than that is a group, which confers lesser advantage). They confer fitness-2 due to group existence in the overall environment (which includes lions in the surroundings), and are selected for by MLS2$_G$. In the case of the bushmen, traits leading to advanced survival prospects include the use of language to communicate with each other, and group hunting of giraffe. These are of course not possible if the group does not exist.

At the individual level, Trait1$_G$ is an individual trait that gives an advantage relative to the environment *because of the existence of the group*; it is a capacity underlying way the individual takes advantage of the existence of the group, which could not occur if the group did not exist. Thus cooperative individuals who are willing to learn will benefit more through the group's existence. Those who ignore group wisdom are likely to perish soon.

The key group level trait underlying all these possibilities is the mere fact of group existence: the tendency to live together as a cooperative group, emergent from the individuals that comprise it. It is the buffalo being together as a herd that allows crucial secondary traits to develop. The same is true *a fortiori* for the bushmen, who through the existence of the group develop far greater survival capacity, particularly through development of technology and tactics that is taught to children.

The key individual level trait underlying the individual benefit accruing from the group, is the individual having a propensity to join a group. They then can benefit by learning from the group, getting its protection, and so on. It is because of this sociable propensity at the individual level that the group emerges from the individuals that comprise it: this is the glue that holds the group together. This confers individual fitness fitness-$1_G$ in regard to the environment because of the existence and nature of the group, and is selected for by $MLS1_G$. Individuals who do not have this propensity to join a group (they like to keep away from the group and go off on their own, are uncooperative and unwilling to learn, and so on) are not so likely to survive, because they do not benefit from the group's existence. This all depends on MLS2, because there would be no advantage to group membership if MLS2 did not exist. Thus the corresponding selective factor reaching down to the individual level is the combination of MLS2 at the group level and $MLS1_G$ in terms of individual behaviour. This depends essentially on emergent group properties (you can't learn from other herd members what is safe to eat and what not, if you don't belong to a herd where this is known). In Section 4, I will consider what mechanisms underlie this tendency to form groups.

In summary: the key point in the present proposal is the idea that one can separate out $MLS1_G$ from $MLS1_E$, and $MLS2_G$ from $MLS2_E$. It seems clear that one can do so in the cases cited here: some group traits confer advantage essentially because of group existence; some related individual traits confer advantage if and only if the group exists. Examples are the capacity to cooperate, and the capacity to learn from others. But the key lower level trait that leads to group level advantage characterised by $MLS2_G$ is the trait of forming meaningful cooperative groups. That is what we focus on below (Section 4). It has played a key role in the evolution of humanity.

**3: LINEAR REGRESSION EQUATIONS**

Linear regression models consider a collective-level trait as part of the context relevant to each particle within that collective, according to the equation (Okasha 2006: 87,198; Pigliucci 2010):

$$w = \beta_1 z + \beta_2 Z + e \tag{1}$$

where w is the particle's fitness, z is the particle-level trait, Z is the collective-level trait, $\beta_1$, $\beta_2$ are regression coefficients, and e is the error term.

According to the present view, one can refine this relation in a number of ways. Define $z_{1E}$ as the particle-level trait solely due to individual fitness, and $z_{1G}$ as the particle-level trait for individual fitness because of group existence. Similarly define $Z_{1E}$ as the group-level trait due to aggregation of individual fitness: $Z_{1E} = \Sigma \, z_{1E}$, and $Z_{1G}$ as the group-level trait essentially due to group existence.

We can then contemplate firstly the group level linear regression equation:

$$W = B_{1E} Z_{1E} + B_{2G} Z_{1G} + e_1 \tag{2}$$

where W is the group's fitness, $B_{1E}$ is the regression coefficient for fitness-1, and $B_{2G}$ the regression coefficients for fitness-$1_G$. Secondly, there is the corresponding particle level equation:

$$w = \beta_{1E} z_{1E} + \beta_{2G} z_{1G} + e_2 \tag{3}$$

where w is the particle's fitness, $\beta_{1E}$ is the regression coefficient for fitness-$1_E$, and $\beta_{2G}$ that for fitness-$1_G$. Finally there are the two multilevel relations:

$$W = B_{1E} \Sigma \, z_{1E} + B_{2G} Z_{1G} + e_{21} \tag{4}$$

expressing the way individual level advantage adds up to group level advantage for aggregate variables, and

$$w = \beta_{1E} z_{1E} + \beta_{12G} z_{1G} Z_{1G} + e_{12} \tag{5}$$

expressing the way group level advantage due to the group existence chains down to individual advantage. The latter equation is non-linear in that it depends on the product $Z := z_{1G} Z_{1G}$, necessary to show how $z_{1G}$ and $Z_{1G}$ act in concert to create selection advantage (see Figure 1); there will be no effect if either of these terms vanishes. It is this non-linearity that underlies the crucial importance of group effects. This non-linearity is because MLS2$_G$ combined with MLS1$_G$ gives a route

$$\text{MLS2} = \text{MLS2}_G \circ \text{MLS1}_G \tag{6}$$

whereby the environment constrains the particle properties in virtue of the combination of selection criteria fitness-2 $_G$ and fitness-$1_G$; they then act up to the emergent level to give the required traits Trait$_2$ at the group level in virtue of the lower level traits Trait$1_G$. The way this happens depends on whether we are concerned with evolutionary (diachronic), developmental, or functional (synchronic) mechanisms and timescales.

The set of equations (2)-(5) obviously allow a more detailed analysis of multilevel effects than does the original equation (1). It is a pragmatic issue as to which of the selection effects represented in (5) by $\beta_{1E}$ and $\beta_{12G}$ is more important in a specific context.

**4: THE UNDERLYING BIOLOGICAL MECHANISMS**

The key issue I now address is the behavioural and biological mechanisms at the individual level that underlies formation and stability of social groups, thereby leading to the group selection advantages **fitness-2** discussed in section 2.

**Behavioural systems:** There are two basic behavioural needs relating the group level to the individual level in order to ensure group formation and stabilisation.

1. **Group formation:** Firstly, in order that meaningful social groups exist, cooperative behaviour between the units that make up the group is crucial: "cooperation amongst lower level units is central to the emergence of new higher levels, because only cooperation can trade fitness from lower to higher levels" (Michod 1999). There are all sorts of mechanisms at the group level intended to make this happen (roles, uniforms, teaching, myths, and so on): "each element making up a social system serves a function that assures the maintenance of the system" (Longres 1990: 39). But that is not enough: in the case of animals and humans, the individual must be responsive to them: they must want to belong to the group. There needs to be an internal mechanism to produce this response.

2.**Group stabilisation**: Generically whenever individuals cooperate together to form groups, there must be mechanisms for regulation of conflict (Okasha 2006: 205), requiring adaptations that suppress within-group competition in order that the group emerge as a genuine whole with adaptations of its own (Okasha 2006: 221-222, 227-228). Again there will be group level mechanisms and processes with this purpose (sanctions, teaching, a legal system, and so on), but by themselves they will not suffice: there must be internal mechanisms to help produce this outcome.

How are these realised at the individual level? In the case of all mammals (including humans), there are plausible biological mechanisms at the individual level underlying both the formation and conflict regulation of such emergent groups. Social groups form because of innate tendencies of individuals to form such groups. The source for this tendency lies in innate primordial emotional systems shared by humans and all higher animals (Panksepp 1998), which are our evolutionary heritage, given to us to help guide decisions we make (Ellis and Toronchuk 2005) and thereby shape intellect (Damasio 1995, 1999). These innate emotional systems function through giving us feelings that tend to produce specific kinds of actions that have promoted survival in our evolutionary ancestors, both human and mammalian (Panksepp and Biven 2012).

In particular there are two such systems that have evolved to create and protect emergent social groups as discussed above. These are (Toronchuk and Ellis 2013),

1. **Group Formation (Trait** $z_{1G}[1]$**)**: an affiliation/attachment system, needed to create such groups, producing feelings of wanting to belong and loneliness when excluded, and starting with mother-child bonding (Stevens and Price 2000:44-45);

2. **Group Stabilisation (Trait** $z_{1G}[2]$**)**: a social ranking system, needed to protect groups by regulating conflict (Stevens and Price 2000:47-48; Price, 1967). This generates a dominance hierarchy which is a social ranking system (the pecking order, the alpha male, etc.); conflict takes place to attain one place in this system, but then acceptance of one's place, and associated territorial rights, regularizes resource allocation in a largely peaceful manner.

These two systems dominate much of social life, and provide the emotional power that enables groups to function. As stated by Stephens and Price (2000:50),

> "*In short, the evidence points to the existence of two great archetypal systems: that concerned with attachment, affiliation, care-giving, care-receiving, and altruism; and that concerned with rank, status, discipline, law and order, territory and possessions. These may well be the basic archetypal patterns on which social adjustment and maladjustment, psychiatric health and sickness depend.*"

These primary emotional systems have evolved over evolutionary times precisely in order to ensure that social groups will come into existence and then be stable (there can be no other reason for their existence as genetically determined systems). We know that they are selected for, because they are innate, and are shared with our ancestral relatives, human and animal (Panksepp and Biven 2012). Thus they are key examples of individual level traits $z1_G$ that have been selected for via individual selection $MLS1_G$, in order to promote group benefits $Z_{1G}$, as indicated by equation (2).

These systems enable the group to come into being, and would not exist as innate systems if there was no major benefit provided by existence of social groups. They have been genetically determined because they are crucial to survival. Putting these together with the kinds of group level traits $Z_{1G}$ discussed in Section 2, the implication is that in equation (5), $z_{1G} Z_{1G} \neq 0$ in both cases.

**Brain Structure:** As in all biology, structure and function are closely linked (Campbell and Reece, 2008). The behavioural mechanisms suggested here must be based in genetically determined underlying neural systems in order to be plausible. This is indeed so: the brain mechanism underlying these behavioural tendencies is the diffusely projecting `ascending systems' of neurons linking nuclei in the limbic system to the neocortex (Kingsley 1996: 132-133), together with the associated neurotransmitters. These systems[3] spread small-molecule neurotransmitters and monoamines such as norepinephrine and dopamine, known to be facilitators of emotions, through the cortex, and hence shape our affective reactions to the world around us (Panksepp 2001, 2012) which are crucial to brain function (Damasio 1995, 1999). They form the `value system' identified by Edelman (1989, 1992) and Edelman and Tononi (2001) as guiding the brain plasticity made possible by neuronal group selection. This is a crucial mechanism for brain plasticity, allowing innate primary emotions to shape adaptation to the physical and social environment in the light of our ongoing experience (Ellis and Toronchuk 2005).

There has to be an evolutionary explanation for the existence of these ascending systems in the brain, which are quite different than the local connections characterising the neural columns that dominate the cortex (Hawkins and Blakeslee 2005). The need for evolution of the innate emotional systems – a key influence in evolutionary history – is the obvious answer. And the need for some of these systems was driven by the evolutionary advantage of group formation, as discussed above.

However not all of these affective systems are of this group-based character. A proposed set of such primary emotional systems is shown in Table 1. This is a summary of the proposals of Toronchuk and Ellis (2013), adapted from those of Panksepp (1998, 2013) by inclusion also of a disgust system (Toronchuk and Ellis 2007) and the ranking system. This proposal is split into two sets: those characterised as fulfilling individual needs (E0-E5), and those fulfilling social needs (E6-E9). The obvious implication is,

> **Multilevel selection proposal:** *(E0-E5) are innate affective systems selected through $MLS1_E$ because of individual advantage given through traits $z_{1E}$, and (E6-E9) are such systems selected through $MLS1_G$ because of group advantage given through traits $Z_{1G}$.*

A case where this may be contentious is Play (E5), which has been classified as an individual trait, even though it plays a considerable role in inter-individual interactions. The view taken here is that the ultimate source of play is indeed in the individual mind, as it explores reality and tries out different options in a creative way – an essential part of the learning process. Group play is derivative from this individual tendency, rather than the other way round.

---

[3] A conveniently accessible paper showing the structure of these systems is Scarr et al (2013).

This proposal accounts for existence both of these behavioural characteristics, and the neural systems (Kingsley 1996: 132-133) that lead to their existence. It has an essentially multi-level nature. If one does not agree with this specific proposal for innate emotional systems, the same issue of the nature of evolutionary origins of the mechanism will arise for any other competing proposal for innate emotional systems (e.g. proposals that there is only one valenced affective system).

**5 THE BIG PICTURE**

This paper has used the specific example of existence of social groups to support the usefulness of the distinction between $MLS1_G$ and $MLS1_E$ (Section 2). Such a distinction is needed to give an evolutionary explanation both for the existence of the emotional systems that are crucial to intellectual functioning (Damasio 1995, 1999), and also for the ascending systems and associated nuclei in the brain that are the neural bases for these emotional systems (Scarr et al 2013). Together with the case of meiotic drive quoted by Okasha (2006), this example serves as an existence proof of causal effectiveness of multilevel selection. If correct, this analysis implies that discussion of evolutionary processes leading to innate behaviour and brain modules should take into account the key role played by emotional systems in the development of these behaviours and modules (Ellis and Toronchuk 2005). Omitting this link, based in specific neuronal mechanisms, can arguably lead to incorrect claims about existence of innate brain modules (Ellis 2008).

The analysis above also confirms the idea of top-down causation in the adaptive selection process (Campbell 1974, Martinez and Moya 2012). Indeed it supports the view that emergence of complexity such as life requires a reverse of flow of information, from bottom-up only to also including a flow from the environment down (Walker, Cisneros and Davies 2012), where "the environment" includes the group. This reverse flow is needed because adaptive selection causes adaptations of the organism to the environment: but this cannot happen unless information flows down from the environment into the organism, where it alters both structure and behaviour. This is what has happened in the past in the case of the coming into being of the innate primary emotional systems, which can be claimed to have played a key role in evolutionary development. However, in contrast to Campbell's compelling example of the jaw of a worker or termite ant (Campbell 1974),[4] the top-down mechanism considered here is essentially multi-level, see Figure 1.

---

[4] This example is considered in depth in Brown and Murphy (2007): 57-58 and in Martinez and Moya (2011): 7/16-8/16.

The analysis presented here emphasizes the importance of taking biological mechanisms seriously in studies of evolutionary selection. The proposal made here suggests further useful developments:

- Seeing how this view extends to the level of cells, where the selection pressures that lead to existence of chemical synapses between neurons, rather than much faster electrical synapses, can plausibly be related to the evolutionary need to develop synaptic plasticity that can be affected by the diffusely projected neuromodulators that form Edelman's `value system' (Edelman 1992).

- Seeing how this view extends even further to the underlying level of genes, where the key feature of multiple realizability, which underlies all top-down causation (Ellis 2012), leads to selection based in equivalence classes of sets of genes, rather than selection of individual genes.

In both these cases (selection at the neuronal level and selection at the genetic level) it is clear that *the selection mechanisms in operation can only be of the multi-level kind discussed in this paper*, as represented in Diagram 1, because there simply is no direct link from the adaptive environment to either neurons or genes. The adaptive causal link in both cases is necessarily via the survival prospects of individual animals on the one hand, and the social groups to which the individuals belong on the other, as discussed above. Multi-level selection, expressed in a suitable extension of Diagram 1, has to be the core of what is going on.

Clearly there will be other contexts where the distinctions proposed in this paper might be useful: indeed they may be relevant anywhere where multilevel selection might perhaps be in play.

∗ ∗ ∗

I thank the NRF (South Africa) for financial support, Trinity College (Cambridge) for a Visiting Commoner Fellowship, and DAMTP (Cambridge) for hospitality in Cambridge.

∗ ∗ ∗

*gfre: 2013-05-16*

\* \* \*

| EVOLUTIONARY NEEDS MET | PRIMARY EMOTIONAL SYSTEM | FUNCTIONS |
|---|---|---|
| *INDIVIDUAL NEEDS* | | |
| Basic Functioning | E0: **PLEASURE** System (satiation, satisfaction) <br> E1 **SEEKING** System | Situation Evaluation, provides arousal/ excitement; incentive salience, hedonic appraisal, and facilitates learning |
| Basic Survival | E2: **DISGUST** System (repulsion) | Avoiding harmful foods, substances, environments |
| | E3: **RAGE** system | Defence: aggression, protection of resources, and con-specifics, |
| | E4: **FEAR** System | Defence: flight, limiting of tissue damage |
| Learning | E5: **PLAY** system | Development of basic adaptive and imaginative skills, creativity [also builds social bonds] |
| *SOCIAL NEEDS* | | |
| Reproduction | E6: **LUST** system <br>    **E6A: Sexual** desire <br>    **E6B: Sexual** satisfaction | Ensuring procreation, enhancement of bonding |
| Group cohesion: Bonding and Development | E7: **PANIC/attachment** (affiliation, separation distress) | Protection of vulnerable individuals; creates bonding through need for others |
| | E8: **CARE** system | Caring for others, particularly offspring |
| Group function: Regulating conflict | E9: **POWER/dominance** system (rank, status, submission) | Controlling aggression in society, allocating resources, esp. sexual ones. |

*Table 1: Evolutionary needs, and the emotional systems that have evolved to meet them.*
E0/E1 is a generalised system providing incentive for the others, and initiated by them. Adapted from Toronchuk and Ellis (2013), which develops from Pankskepp (1998).